\documentstyle[12pt,aasms4,psfig]{article}

\begin{document}
\title{\bf Dark energy and the epoch of galaxy formation}
\author{J.S.Alcaniz\altaffilmark{1} and J.A.S.Lima\altaffilmark{2}}
\affil{Departamento de F\'{\i}sica, Universidade Federal do Rio Grande do 
Norte, 
\\ 
C.P. 1641, 59072-970, Natal, Brazil}
\altaffiltext{1}{alcaniz@dfte.ufrn.br}
\altaffiltext{2}{limajas@dfte.ufrn.br}

\vspace{1.0cm}

\centerline{\it Accepted for publication in ApJL}

\begin{abstract}
The influence of a dark  component on the first epoch of galaxy
formation  is analysed by using the ages of the three
oldest high-redshift galaxies known in the literature. Our results, based on a
spatially  flat accelerated Universe driven by a ``quintessence" component
($p_x = \omega\rho_x$), show that if the inferred ages of these objects are
correct the first formation era is pushed back to extremely high redshifts. 
For the present best-fit quintessence model ($\Omega_{x}=
0.7$, $\omega < -0.6$), we find a lower bound of $z_f \geq 7.7$, whereas
in the extreme case of $\Lambda$CDM model ($\omega = -1$) the limit  is
slightly smaller ($z_f \geq 5.8$). The case for open cold dark matter models
(OCDM) has also been discussed. For $\Omega_m \simeq 0.3$, the formation
redshift is restricted by $z_f \geq 18$. As a general result, if $\Omega_m
\geq 0.37$, these galaxies are not formed in FRW cosmologies with no dark
energy since for all these cases $z_f \rightarrow \infty$. 
\end{abstract}

\pagebreak

In the last decade, the remarkable observational progress in addressing the 
question of the origin,  early evolution and ages of galaxies, as well as the
expected rapid improvement of these data coming from the next generation of
space telescopes may provide the key for a precise determination of the epoch
when the most of galaxies in the Universe were formed. 

Recently, different groups (Stockton {\it et al} 1995; Dunlop {\it et al}
1996; Spinrard {\it et al} 1997; Dunlop 1998) announced the discoveries of
three extremely red radio galaxies at $z=1.175$ (3C 65), $z=1.55$ (LBDS
53W091) and $z=1.43$ (LBDS 53W069) with a minimal  stellar age of 4.0 Gyr, 3.5 
Gyr and 4.0 Gyr, respectively. These discoveries accentuated even further the
already classical ``age crisis'' and  gave  rise to a new variant of this
problem, which could be named the high-$z$ time scale crisis: the
underestimated ages of these galaxies  contradict the predictions of the
standard Einstein-de Sitter model for values of  $h < 0.45$ (Krauss 1997).
Furthermore, for $\Lambda$CDM scenarios, the age-redshift relation for these
OHRGs requires $\Omega_{\Lambda} \geq 0.50$ if the Universe is assumed to be
spatially flat (Alcaniz \& Lima 1999; Lima \& Alcaniz 2000a). The problem
raised by the ages of these galaxies to the standard CDM model suggests that
OHRGs  may be an important piece of the missing information related to the
galaxy formation process, and in a similar vein, may constrain the epoch when
the first structures were actually formed.

On the other hand, recent measurements of some Type Ia Supernovae at
intermediate  and high redshifts (Riess {\it et al} 1998; Perlmutter {\it et
al} 1999) also indicate that the bulk of energy in the Universe is repulsive
and appears like a ``quintessence" component, that is, an unknown form of dark
energy  (in addition to the ordinary CDM matter) probably of primordial origin
(see also Turner 2000 for a review). Together with the observations of CMB
anisotropies (de Bernardis 2000), these results provide the remaining piece
of information connecting the inflationary flatness prediction
($\Omega_{\rm{T}} = 1$) with the astronomical observations. This state of
affairs has  stimulated the interest for more general models containing an
extra component describing this dark energy, and simultaneously accounting for
the present accelerated stage of the Universe. In the last years, the absence
of a convincing evidence on the nature of the dark component gave origin to an
intense debate and mainly to theoretical speculations. Some possible
candidates for ``quintessence" are: a vacuum decaying energy density, or a
time varying $\Lambda$-term (Ozer \& Taha 1987; Freese 1987; Carvalho {\it et
al} 1992), a relic scalar field (Peebles \& Ratra 1988; Frieman {\it et al}
1995; Caldwell {\it et al} 1998; Saini {\it et al} 2000) or still an extra
component, the so-called ``X-matter", which is simply characterized by an
equation of state $p_x=\omega\rho_{x}$, where $\omega\geq -1$ (Turner \& White
1997; Chiba {\it et al} 1995; Lima \& Alcaniz 2000b) and includes, as a
particular case, models with a cosmological constant ($\Lambda$CDM) (Turner
{\it et al} 1984; Peebles 1984; Krauss \& Turner 1995). For ``X-matter" models,
constraints from large scale structure (LSS) and cosmic microwave background
anisotropies (CMB) complemented by the SN Ia data, require $0.6 \leq \Omega_x
\leq 0.7$ and $\omega < -0.6$ ($95\%$ C.L.) for a flat universe (Perlmutter
{\it et al} 1999; Efstathiou 1999; Garnavich {\it et al} 1998), while for
universes with arbitrary spatial curvature the limit is $\omega < -0.4$
(Efstathiou 1999). 

In this {\it letter} we focus our attention on this kind of ``quintessence'' 
or X-matter cosmology.  As is widely known, because of their generality these 
models also merit a broader discussion. In principle, to check the validity
of a theory or model (for instance, $\Lambda$CDM model), it is interesting to
insert it in a more general framework, herein quantified by the $\omega$
parameter. In this context, we derive new lower limits to the formation
redshift $z_f$ from ages of the oldest high redshift objects known in the
literature. The main consequences of this approach for the standard open FRW
model (OCDM) are also briefly analysed.

For passively evolving elliptic radio galaxies, almost all amount  of gas is
believed  to be processed into stars in a single episode of star formation,
in such a way that the assumption of an instantaneous burst is considered a
good approximation for modeling their evolution (Jimenez {\it et al} 1999). For
lookback time calculations, the hypothesis of an instantaneous burst of star 
formation means that the age of these OHRGs can be expressed as being almost
exactly the time taken by the Universe to evolve from $z_f$ to the observed
redshift $z_{obs}$. In the framework of  flat Friedmann-Robertson-Walker (FRW)
models with nonrelativistic matter plus a quintessence component ($p_{x} =
\omega\rho_{x}$, $\omega \geq -1$) such condition can be translated as 
\begin{eqnarray}  
t_{z_{obs}} - t_{z_{f}} & = &
H_{o}^{-1}\int_{(1+z_f)^{-1}}^{(1 + z_{obs})^{-1}}  {dx \over
x\sqrt{\Omega_{x}x^{-3} +   \Omega_{x} x^{-3(1 + \omega)}}}  \nonumber \\  &  
& \geq t_g  \quad,  
\end{eqnarray}   
where $\Omega_x$ stands for the
present-day quintessence density parameter. The inequality signal on the
r.h.s. of the above expression comes from the fact that the Universe is older
than or at least has the same age of any observed structure. Since this
natural argument also holds for any time interval, a finite value for the
redshift $z_f$ provides the lower bound for the galaxy  formation allowed by
the aged object located at $z_{obs}$. Models for which $z_f \rightarrow
\infty$ are clearly incompatible with the existence of the specific galaxy,
being ruled out in a natural way. Indeed, as argued by Peebles (1989), any
model is already in trouble if $z_f$ is bigger than the finite redshift for
which the mean density of the Universe was equal to the mean density of the
aged high redshift galaxy (null contrast density).  

Before discussing the resulting diagrams, an important  point of principle
should be stressed. To assure the robustness of the limits on $z_f$, we addopt
(1) the minimal value for the Hubble parameter,  and (2) the underestimated age
for all OHRGs. Both conditions are almost
self-explanatory. First, as we know, the smaller the value of $H_{o}$, the
larger the age predicted by the model and, second, objects with smaller
ages are theoretically more easily accommodated, thereby guaranteeing that the models are
always favored in the present estimates. For the Hubble parameter we consider the 
value obtained by the HST Key project which is in agreement with other
independent estimates (Giovanelli {\it et al} 1997), i.e., the  round
number value $H_{o} = 60\rm{km/sec/Mpc}$ (Freedman 1998). Indeed, we are being
rather conservative since this lower limit was recently updated to nearly
$10\%$ of accuracy ($h = 0.70 \pm 0.07$, $1\sigma$) by Freedman (2000), 
and the data from SNe also point consistently to $h > 0.6$ or even higher
(Riess {\it et al} 1998; Perlmutter {\it et al} 1999). 

At this point, some few words concerning the ages of these galaxies and their 
consequences are also appropriate.  Since its discovery, the age estimates
for the  LBDS 53W091 (at $z = 1.55$) has stimulated considerable debate and
controversy in the literature. Initially,  by analyzing the strong transition
breaks existing in the spectrum at 2640${\rm{\AA}}$ and 2900${\rm{\AA}}$,
Dunlop {\it et al.} (1996) and  Spinrad {\it et al} (1997) found
a best fitting age of 3.5 Gyr.  Later on, Bruzual \& Magris (1997) and 
Yi {\it et al} (2000), using different arguments, claimed to have derived 
an age as young as 1.5 Gyr. More recently, we witnessed a new  chapter on this
controversy. Through a more detailed analysis of the different evolution 
models (based on $\chi^{2}$ minimization), the minimum age of 3.0 Gyr has
again been clearly favored (Nolan {\it et al} 2001). For the radio galaxy 3C
65 at $z = 1.175$, we adopt the estimated age from the analysis by Stockton
{\it et al.} (1995) which has indicated strong evidence for a minimum
stellar age of 4.0 Gyr. However, as we shall see, since the redshift of this
galaxy is comparatively low, the corresponding constraint on the first epoch of
formation is the less restrictive one. Finally, the remaining galaxy (LBDS
53W069, $z=1.43$) is an even redder galaxy which is 1.0 Gyr older than 53W091,
and whose inferred age seems to be essentially unchanged for different solar
metallicity models (Nolan {\it et al} 2001). Actually, for a given value of
$\Omega_m$, the tighter constraints on $z_f$ come from this object.  

In Figure 1 we show the $\omega - z_f$ plane allowed by the  existence of
these OHRG's for quintessence models.  In order to have a bidimensional plot
one needs to fix the value of one of the parameters ($\Omega_x$ or $\omega$). 
Here we use the observational limits from LSS+CMB+SNe Ia, i.e., $\Omega_x
\simeq 0.7$ and $\omega < -0.6$  at $95\%$ c.l. (Perlmutter {\it et al}
1999; Efstathiou 1999; see also Wang {\it et al} 2000 for limits from
``concordance cosmic" method). This constraint is represented by the shadowed
horizontal region and is used to determine the lower limits on $z_f$. Since
the effect of the equation of state associated to the quintessence component
($\omega = p_x/\rho_x$) is to  accelerate the cosmic expansion,  this means
that the lookback time between $z_{obs}$ and $z_f$ is larger than in the
standard scenario and, therefore, the galaxy formation process may start
relatively late in comparison to the corresponding  standard CDM model. For
example, for the interval considered here ($-0.6 \leq \omega < -1$) the
estimated age of the 3C 65, 53W091 and 53W069 provides, respectively,  $z_f
\geq 4.5 - 3.6$, $z_f \geq 6.4 - 5.2$ and $z_f \geq 7.7 - 5.8$, with the lower
values of $z_f$  corresponding to the lower $\omega$. These lower bounds are
also heavily dependent on the Hubble parameter. For instance, on the interval
$0.6 \leq h  \leq 0.7$, the formation redshift for the best-fit quintessence
model ($\Omega_{x}= 0.7$, $\omega < -0.6$) (Turner \& White 1997;
Perlmutter {\it et al.} 1999; Efstathiou 1999; Garnavich {\it et al.} 1998) 
varies from $z_f \geq 7.7$ to $z_f \geq 30$ when the age of the LBDS 53W069 is
considered. Naturally, if the inferred ages of these OGHRs were smaller, the
corresponding formation redshifts should be relatively smaller (for fixed
values of the other parameters) in comparison with the limits derived.  For
example, by taking the estimated age for the LBDS 53W091 given by Bruzual \&
Magris (1997) and Yi {\it et al} (2000), i.e., $t_g \sim 1.5$ Gyr, the
formation redshift is reduced from $z_f \geq 6.4$ to $z_f \geq 2.3$ for the
interval of parameters considered above. However, if the age of this object is
greater than 2.5 Gyr (J. S. Dunlop, private communication) the corresponding
lower bound is $z_f \geq 2.9$.

Figures 2 and 3 display the $\Omega_m - z_f$ plane  for OCDM and
$\Lambda$CDM models, respectively. The shadowed horizontal region corresponds
to the observed range $\Omega_{\rm{m}} = 0.2 - 0.4$ (Dekel {\it et al.} 1997),
which is also used to fix the lower limits on $z_f$. As should be  physically
expected, in the case of OCDM models (Fig. 2), if the matter contribution
increases, a larger value of $z_f$ is  required in order to account for the
existence of these OHRGs within these models. Conversely, for each object, the
absolute minimal value of $z_f$ is obtained for an empty universe ($\Omega_m
\rightarrow 0$). In the observed range of $\Omega_m$ the allowed values for
the formation redshift are unexpectedly high. 
For example, by considering $\Omega_{\rm{m}} = 0.3$, as indicated from 
dynamic estimates on scales up to about 2$h^{-1}$ Mpc (Calberg {\it et al}
1996), the ages of 3C 65, 53W091 and 53W069 provide, respectively, $z_f \geq
6.3$, $z_f \geq 10.5$ and $z_f \geq 18$. Such values suggest that these
galaxies were formed about 12.5 Gyr ago, or by considering the most recent
lower limits for the age of  the Universe (Carretta {\it et al} 2000; Krauss
2000), that such objects were formed when the Universe was $\sim$ 1.0 Gyr old.
However, since almost all the age of the Universe is at low redshifts ($z = 0
- 2$), these galaxies may have been formed nearly at the same epoch,
regardless of their constraints on the redshift space. It is also interesting
to note that the value of $z_f$ is proportional to  the quantity of dark
matter. As one may check, in the limiting case of a universe dominated only by
dark energy ($\Omega_{\rm{m}} = 0$, $\omega = -1$), the lower limit is $z_f
\geq$ 2 (see Fig. 3).

Tables 1 and 2 summarize the estimated values of $z_f$ for quintessence and
CDM  models. In the former case the
matter density parameter has been fixed in its central observed value
($\Omega_m = 0.3$), with $z_f$ determined for several values of $\omega$
(see Table 1). As explained before, since the lookback time between $z_f$ and
$z_{obs}$ increases for lower values of $\omega$, it is physically expected
that the value of $z_f$ diminishes for smaller values of $\omega$. The
remaining cases, OCDM and $\Lambda$CDM, are displayed for the entire observed
range of the matter density parameter ($\Omega_{\rm{m}}$ = 0.2 - 0.4).  Note
that for $\Omega_{\rm{m}} \geq 0.37$, the lower limit inferred from the age of
59W069 is $z_f \rightarrow \infty$. This result is consistent with recent
studies based on the age-redshift relation (Alcaniz \& Lima 1999), and means
that the standard cosmological model  with $\Omega_{\rm{m}} \geq 0.37$ and  $h
\geq 0.6$ is (beyond doubt)  incompatible with the existence of this galaxy.
Our results are also in line with the claims of Kashlinsky and Jimenez (1997) 
that the galaxy 53W091 represents a very rare and unlikely event in the density
field of such models. In principle, they also imply that quintessence models
contitutes a potential alternative to conciliate the observed excess of  power
observed in galaxy surveys without requiring too late galaxy formation (as it
happens in $\Lambda$CDM models). It should be
stressed that the present constraints on $z_f$ are indeed rather conservative
since the lower limit on $H_o$ has been considered in all the estimates.
Furthermore, the inferred ages for these galaxies refers to the end of the
star formation episode that dominates the spectrum and not to the first epoch
of star formation in the galaxy what, therefore, provides only a lower limit
to the age of the galaxy. Finally, these high limits on $z_f$ provide a new
theoretical evidence that galaxies are not uncommum objects at very large
redshifts, say, at $z > 5$, and also reinforce the interest on the
observational search for galaxies and other collapsed objects within the
redshift interval ($5 \leq z \leq 10$), which nowadays delimits the last
frontier to a more complete understanding on the galaxy formation process.  

\acknowledgments
\section*{Acknowledgments}

The authors are grateful to Robert Brandenberger, Am\^ancio Fria\c{c}a and
Sueli Viegas for  helpful discussions. This work was  partially suported by
the Conselho Nacional de Desenvolvimento Cient\'{\i}fico e  Tecnol\'{o}gico
(CNPq), Pronex/FINEP (no. 41.96.0908.00) and FAPESP (no. 00/06695-0).


\newpage

\pagestyle{empty}

\begin{center}
\begin{table*}   
\centering   
\begin{center} 
\caption{Limits to $z_f$: Quintessence models} 
\end{center} \begin{tabular}{rll}  
\hline  
\multicolumn{1}{c}{Galaxy}& 
\multicolumn{1}{c}{$\omega$}&
\multicolumn{1}{c}{$z_f$}\\
\hline 
3C 65:...........& -1/3  & $\geq$ 6.6\\   
 & -0.6 & $\geq$ 4.5\\ 
& -0.8 & $\geq$ 3.7\\ 
53W091:........& -1/3 & $\geq$ 11.2\\   
 & -0.6 & $\geq$ 6.4\\ 
& -0.8 & $\geq$ 5.5\\ 
53W069:........& -1/3 & $\geq$ 20.3\\   
 & -0.6 & $\geq$ 7.7\\ 
& -0.8 & $\geq$ 6.3\\ 
\hline 
\end{tabular} 
\end{table*} 
\end{center}

\begin{center} 
\begin{table*}  
\centering 
\begin{center} 
\caption{Limits to $z_f$: OCDM and $\Lambda$CDM models} 
\end{center} \begin{tabular}{rllr}  
\hline  \multicolumn{1}{c}{Galaxy}& 
\multicolumn{1}{c}{$\Omega_{\rm{m}}$}&
\multicolumn{1}{c}{$\frac{\rm{OCDM}}{z_{f }}$}&
\multicolumn{1}{c}{$\frac{\rm{\Lambda CDM}}{z_{f}}$}\\ 
\hline 
3C 65:...........&0.2 & $\geq 5.0$& $\geq
3.0$\\   
 & 0.3 & $\geq 6.3$& $\geq 3.6$\\ 
&0.4& $\geq 8.6$& $\geq 4.8$\\ 
53W091:........& 0.2 & $\geq 7.4$& $\geq
3.8$\\   
 &0.3& $\geq 10.5$&  $\geq 5.2$\\ 
&0.4& $\geq 20.7$& $\geq 7.4$\\ 
53W069:........&0.2 & $\geq 9.3$& $\geq
3.9$\\   
 &0.3& $\geq 18.0$& $\geq 5.8$\\ 
&0.37& $\rightarrow \infty$& $\geq 10$\\ 
\hline 
\end{tabular} 
\end{table*} 
\end{center}

\newpage

\begin{figure} 
\vspace{.2in}  
\centerline{\psfig{figure=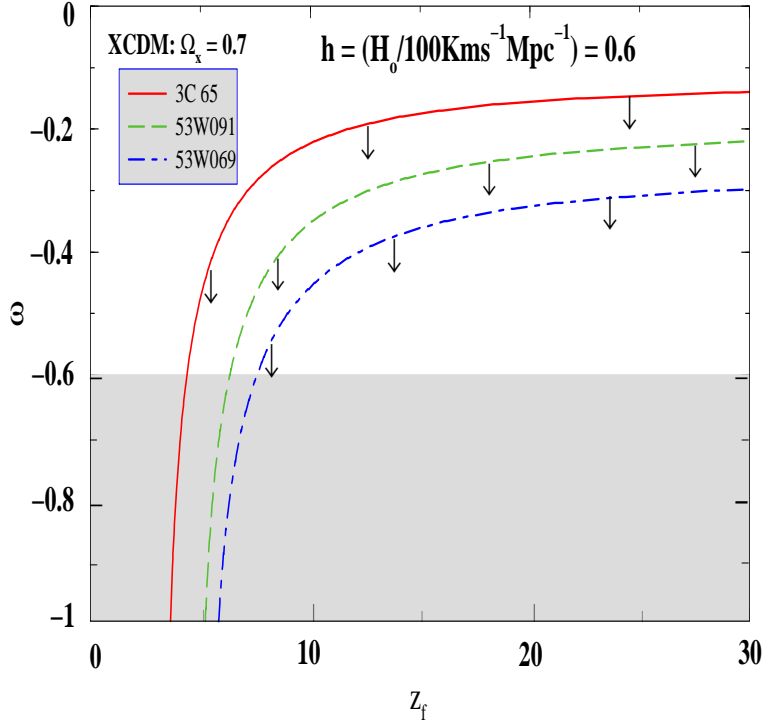,width=4.0truein,height=4.0truein}  
\hskip 0.1in}   
\caption{The $\omega - z_f$ plane for quintessence models allowed by the
existence of OHRG's. The value of $\Omega_x$ is indicated in the figure and
the shadowed region corresponds to the range of $\omega$ as determined from
LLS+CMB+SNe Ia analysis (Perlmutter {\it et al.} 1999; Efstathiou 1999;
Garnavich {\it et al.} 1998). As explained in the text, for each high redshift
object, the arrows delimit the available parameter space. The curves are
defined by the underestimated values of $t_g$ and the indicated lower limit of
$H_o$. For the best fit model ($\Omega_x = 0.7$, $\omega < -0.6$), the
corresponding lower limit is $z_f \geq 7.7$ (see also Table 1).}
 \end{figure}

\begin{figure} 
\vspace{.2in}  
\centerline{\psfig{figure=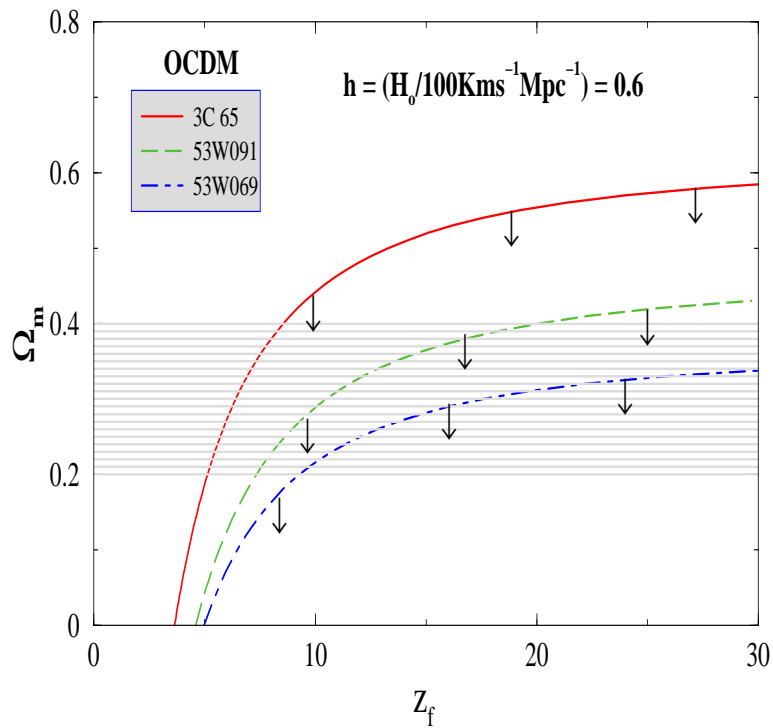,width=4.0truein,height=3.9truein}  
\hskip 0.1in}   
\caption{The $\Omega_m - z_f$ plane allowed by the  existence of OHRG's in 
the framework of open cold dark matter models (OCDM). The shadowed
horizontal region corresponds to the observed range of $\Omega_{\rm{m}}$. 
The arrows delimit the available parameter space. The curves are also defined 
by the underestimated values of $t_g$ and the indicated lower limit of $H_o$.
For a given value of $\Omega_m$, we see that the most restrictive limit is
provided by the radio galaxy 53W069.} 
\end{figure}   

\begin{figure} 
\vspace{.2in}  
\centerline{\psfig{figure=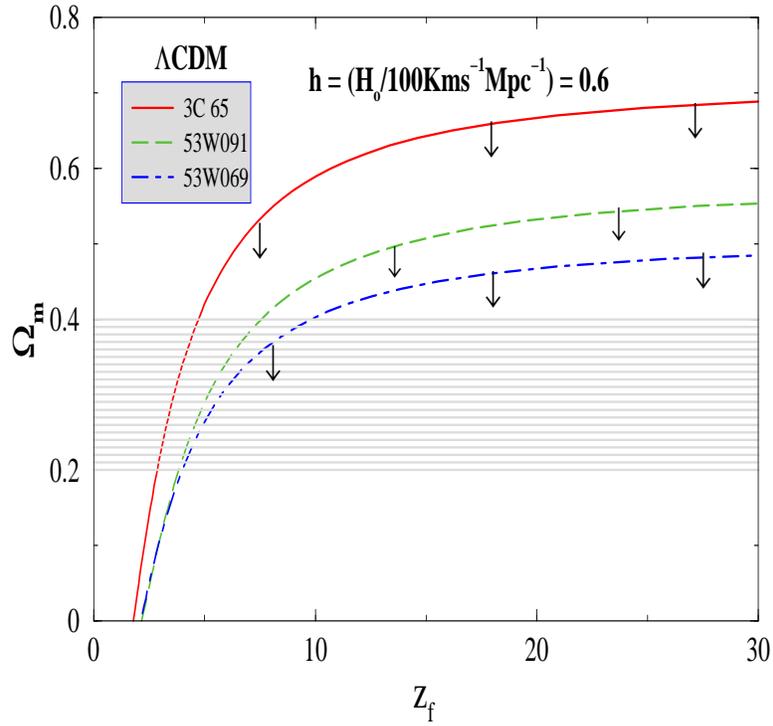,width=4.0truein,height=3.9truein}  
\hskip 0.1in}   
\caption{The same plot as in Fig. 2 for $\Lambda$CDM models. Again, the contourns are 
obtained by considering the minimal value of $t_g$, and the
shadowed horizontal region corresponds to the observed range of 
$\Omega_{\rm{m}}$. In comparison to the corresponding OCDM model, we see that 
all curves have been shifted for smaller redshifts (see also Table 2).}
\end{figure}

\end{document}